\documentclass[10pt,prl,aps,twocolumn,showpacs,
superscriptaddress]{revtex4}
\usepackage{graphicx}

\begin{document}

\title{Comment on ``Ground State Phase Diagram of a Half-Filled
One-Dimensional Extended Hubbard Model''}

\author{Anders W. Sandvik} 
\affiliation{Department of Physics, {\AA}bo Akademi University, 
Porthansgatan 3, FIN-20500, Turku, Finland}

\author{Pinaki Sengupta} 
\affiliation{Department of Physics, University of California,
Davis, California 95616}

\author{David K. Campbell}
\affiliation{Departments of Physics and of Electrical and Computer 
Engineering, Boston University, 44 Cummington Street, Boston, 
Massachusetts 02215}

\date{August 20, 2003}


\maketitle

In \cite{jeckelmann}, Jeckelmann argued that the recently discovered
bond-order-wave (BOW) phase \cite{nakamura,pinaki,tf} of the 1D extended 
Hubbard model does not have a finite extent in the $(U,V)$ plane, but exists 
only on a segment of a first-order spin-density-wave--charge-density-wave
(SDW-CDW) phase boundary. We here present 
quantum Monte Carlo results of higher precision and for larger system sizes 
than previously \cite{pinaki}. Using a direct finite-size scaling of the
BOW correlations, we reconfirm that the BOW phase {\it does exist} a finite 
distance away from the phase boundary, which hence is a BOW-CDW transition 
curve. We only address the existence of the BOW phase and focus on a single 
value, $U=4$, of the on-site interaction.

\begin{figure}
\includegraphics[width=5.6cm]{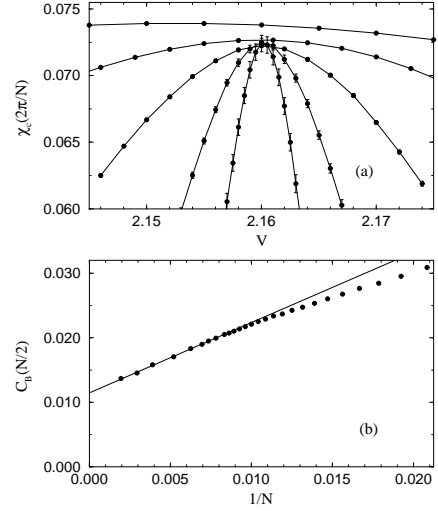}
\caption{(a) Charge susceptibility at $q=2\pi/N$ for $N=32,64,128,256$, 
and $512$ (curvature increases with $N$). (b) BOW correlation at distance 
$r=N/2$ versus inverse system size ($V=2.10$), along with a linear fit to 
the large-$N$ data.}
\label{fig}
\end{figure}

We first determine the critical value $V_c$ of the nearest-neighbor
interaction. Fig.~\ref{fig}(a) shows the $V$-dependence of the charge 
susceptibility $\chi_c(q)$ at the smallest non-zero wave-number, $q=2\pi/N$, 
for different system sizes $N$ (for the definition of $\chi_c$, see 
\cite{pinaki}). The narrowing of the peak with increasing $N$ and the 
convergence of the height indicate a charge gap for all $V$ except at $V_c$, 
i.e., in disagreement with \cite{jeckelmann} we find a continuous phase 
transition at $U=4$. The peak position gives $V_c=2.1602 \pm 0.0003$, which 
improves considerably on the estimate $V_c=2.150 \pm 0.010$ reported
in \cite{jeckelmann}.

We next choose $V=2.10$, where according to \cite{jeckelmann} there should
be no long-range BOW. In Fig.~\ref{fig}(b) we show the corresponding
correlation function $C_B(r)$ at the longest distance, $r=N/2$, in periodic
chains with up to $N=512$ sites. As a function of $1/N$ for large $N$, the 
data scale linearly to a value which corresponds to dimerization $\delta=
\sqrt{C_B}/2\approx 0.053$ (as defined in \cite{jeckelmann}). It is not 
clear why the density matrix renormalization group (DMRG) 
calculations of \cite{jeckelmann} failed to detect this 
rather strong BOW order. Dimerization was observed only on the transition 
curve to the CDW phase \cite{jeckelmann}, where we instead find coexisting
critical BOW and CDW fluctuations. The origin of this discrepancy at $V_c$ 
could be that $V_c$ was not determined to sufficient accuracy in 
\cite{jeckelmann}. 

In summary, our improved calculations do not agree with the phase diagram 
presented in \cite{jeckelmann} but reconfirm the finite extent of the BOW 
phase \cite{pinaki} as first suggested in \cite{nakamura}. The advantage of 
the Monte Carlo method we have used \cite{pinaki,sse} is that results can be 
obtained for large periodic systems, which are better suited for finite-size 
scaling than the open boundary conditions typically used in DMRG calculations 
on long chains. With the recently improved simulation algorithm \cite{sse} 
that we have used here, we hope to be able to determine the full extent of 
the BOW phase in the $(U,V)$ plane.

This work is supported by the Academy of Finland, project 26175 (AWS),
and by the NSF under Grants No.~DMR-99-86948 (PS) and DMR-97-12765 (DKC). 
The calculations were carried out at the NCSA in Urbana, Illinois, and at 
the CSC in Helsinki.

\null

\end{document}